\documentclass[aps, pra, superscriptaddress, twocolumn, amsfonts, amsmath, amssymb, floatfix]{revtex4-2}
\usepackage{graphicx}
\usepackage{float}
\usepackage{sidecap}
\usepackage{dcolumn}
\usepackage{bm}
\usepackage{epsfig}
\usepackage{braket}
\usepackage{color}
\usepackage{ulem}
\usepackage{amsmath}
\usepackage[colorlinks=true, pdfstartview=FitV, linkcolor=blue, citecolor=blue, urlcolor=blue]{hyperref}

\begin{document}

\title{Expansion dynamics of Bose-Einstein condensates in a synthetic magnetic field}

\author{Juan Wang}
\affiliation{Department of Physics, School of Physics and Electronic Science, East China Normal University, Shanghai 200241, China}

\author{Hongguang Liang}
\affiliation{Department of Physics, School of Physics and Electronic Science, East China Normal University, Shanghai 200241, China}

\author{Yan Li}
\email{yli@phy.ecnu.edu.cn}
\affiliation{Department of Physics, School of Physics and Electronic Science, East China Normal University, Shanghai 200241, China}

\author{Chuan-Hsun Li}
\affiliation{Department of Physics and Astronomy, Purdue University, West Lafayette, Indiana 47907, USA}

\author{Chunlei Qu}
\email{cqu5@stevens.edu}
\affiliation{Department of Physics, Stevens Institute of Technology, Hoboken, NJ 07030, USA}
\affiliation{Center for Quantum Science and Engineering, Stevens Institute of Technology, Hoboken, NJ 07030, USA}

\begin{abstract}
We investigate the expansion dynamics of spin-orbit-coupled Bose-Einstein condensates subjected to a synthetic magnetic field, after their release from an external harmonic trap. Our findings reveal that the condensate experiences a spin-dependent rotation and separation due to the rigid-like rotational velocity field, which leads to a spin density deflection. The deflection angle reaches a peak at a time that is inversely related to the frequency of the harmonic trap. When the detuning gradient is below a critical value for vortex nucleation, our analytical results derived from a spinor hydrodynamic theory align closely with numerical results using the coupled Gross-Pitaevskii equations. Beyond this critical value, we also numerically simulated the expansion dynamics of the condensates containing vortices with negative circulation. Our findings highlight the pivotal role of the rigid-like rotational velocity field on the dynamics of the condensate and may stimulate further experimental investigations into the rich superfluid dynamics induced by synthetic magnetic fields.
\end{abstract}

\maketitle

%\tableofcontents

\section{Introduction}
The investigation of expansion dynamics of quantum gases released from external traps has significantly advanced our understanding of their quantum states and the roles of interactions \cite{PhysRevLett.89.250402}. In 1995, the detection of the anisotropic velocity distribution in an atomic cloud after expansion provided pivotal evidence for the existence of Bose-Einstein condensates (BECs) \cite{doi:10.1126/science.269.5221.198,PhysRevLett.75.3969}. Since then, the measurement of expansion dynamics has become a powerful tool for probing a plethora of physical phenomena in both bosonic and fermionic quantum gases \cite{PhysRevLett.80.2027,lin2002quantum,EurophysLett.10.1209/epl/i1998-00107-2,PhysRevLett.92.230402,lin2008anderson,WANG20207}. For example, the observation of quantized vortices in BECs released from rotating traps has revealed the superfluid characteristic of Bose gases \cite{PhysRevLett.83.2498,PhysRevLett.84.806}. 

In recent years, the experimental and theoretical exploration of synthetic gauge fields in ultracold atomic gases have made rapid progress and opened new frontiers in modern quantum physics \cite{PhysRevA.79.063613,PhysRevLett.102.130401,lin2009synthetic,PhysRevLett.105.160403,lin2011spin,PhysRevLett.107.150403,PhysRevLett.107.200401,PhysRevA.84.063604,PhysRevLett.109.115301,PhysRevLett.108.010402,PhysRevLett.108.225301,PhysRevLett.108.035302,galitski2013spin,PhysRevA.87.063610,PhysRevA.94.033635,zhang2016properties}. The implementation of Raman coupling via couter-propagating laser beams to couple two hyperfine states of alkaline atoms has enabled the realization of an equal superposition of Rashba and Dresslhaus spin-orbit (SO) coupling \cite{lin2011spin}. SO coupling significantly affects the superfluidity of Bose condensates, including the quenching of the superfluid density along the SO coupling direction near the critical transition point between the zero-momentum phase and plane-wave phase \cite{PhysRevA.86.063621,PhysRevA.94.033635}. Additionally, introducing a position-dependent detuning to the SO-coupled Hamiltonian enables the generation of a synthetic magnetic field for the neutral atoms \cite{lin2009synthetic}. This has led to the observation of quantized vortices without a rotating trap \cite{PhysRevA.84.063604}. Even before the appearance of vortices, the synthetic magnetic field induces a rigid-like rotational velocity field, violating the irrotational superfluid velocity field constraint for traditional condensates \cite{PhysRevLett.118.145302}. Synthetic magnetic fields have also facilitated novel investigations into the non-equilibrium dynamics of SO-coupled BECs. Recently, it was found that a synthetic magnetic field couples various low-energy collective modes of the condensate, resulting in intriguing beat phenomena \cite{PhysRevLett.120.183202,PhysRevA.108.053316}.

\begin{figure}[t!]
\centering
\includegraphics[width=1\linewidth]{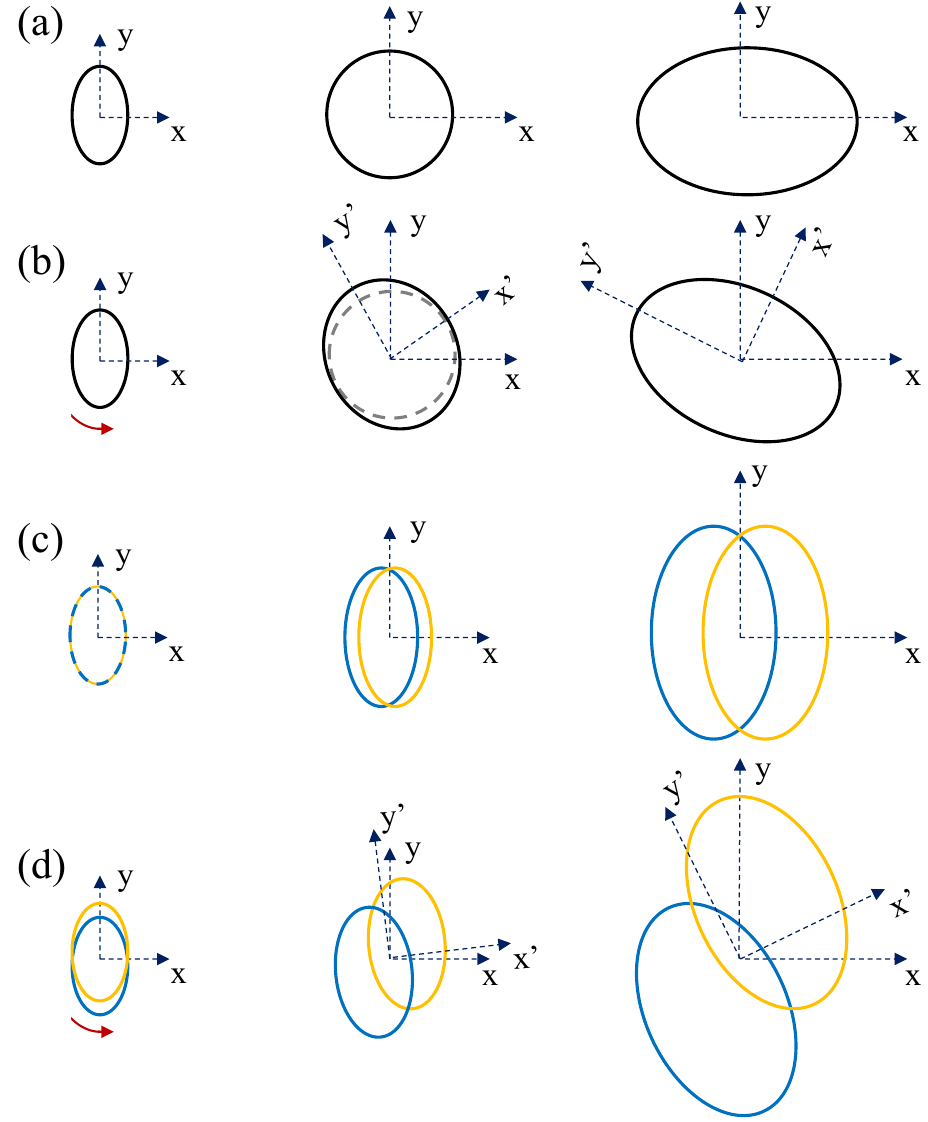}
\caption{Schematic of the expansion dynamics of different types of condensates. From left to right, the three columns illustrate the initial state, short-time expansion, and long-time expansion, respectively. (a) An ordinary BEC shows an inversion of the aspect ratio. (b) A rotating BEC never becomes circular (indicated by the dotted circle) during the expansion. (c) A SO-coupled BEC expands significantly slower along the SO direction, with the two spin components (yellow and blue lines) separating as they expand. (d) A BEC in a synthetic rotational field exhibits both rotation and separation during the expansion. For (c) and (d), the SO coupling is along $x$. The red arrows in (b) and (d) indicate the direction of the rotating trap or rotational velocity field.}
\label{fig:bec}
\end{figure}

In this work, we study the expansion dynamics of the condensate in the presence of a synthetic magnetic field. Figure~\ref{fig:bec} illustrates the key characteristics of the expansion dynamics across various condensates. For a non-rotating, single-component BEC, an inversion of the aspect ratio is observed during the expansion (Fig.~\ref{fig:bec}(a)). This is attributed to the mean-field interaction energy and the phenomenon can be accurately described by the hydrodynamic theory \cite{PhysRevLett.77.5315,PhysRevA.54.R1753,PhysLettA.10.1016/S0375-9601(97)00069-8}. In contrast, as shown in Fig.~\ref{fig:bec}(b), a rotating, vortex-free condensate does not exhibit the inversion of aspect ratio due to the irrotational superfluid flow and the conservation of angular momentum \cite{PhysRevLett.88.070405,PhysRevLett.88.070406}. Figure~\ref{fig:bec}(c) depicts the expansion of a SO-coupled BEC which is characterized by a significantly slower expansion along the SO coupling direction $x$, due to a large effective mass~\cite{NewJphys.10.1088/1367-2630/aa7e8c}. Moreover, the spin-momentum locking dictates that the two spin components (shown by yellow and blue curves) separate along the SO coupling direction during the expansion, creating a spin density along $x$~\cite{PhysRevLett.108.035302}. 

The expansion dynamics of a condensate in a synthetic magnetic field exhibit a different behavior. As shown in Fig.~\ref{fig:bec}(d), the two spin components exhibit both rotation and separation in the expansion, leading to a deflection of the spin density that is intimately related to the rigid-like rotational velocity field. This deflection dynamics can be analytically determined by solving the hydrodynamic equations for a vortex-free BEC in a weak synthetic magnetic field. Moreover, when the synthetic magnetic field is sufficiently strong, by solving the coupled Gross-Pitaevskii (GP) equations, we find that the expansion dynamics exhibit an additional rotation of the vortices.

\section{Hydrodynamic theory} 
We consider the following single-particle Hamiltonian for a BEC in a synthetic magnetic field \cite{PhysRevA.79.063613,lin2009synthetic}:
\begin{align}
\hat{H}_{sp} & = \frac{1}{2m}( \hat{p}_{x} -\hbar k_{r} \sigma _{z}) ^{2}+ \frac{\hat{p}_{y}^{2}}{2m}+  V_{trap} -\frac{\Omega }{2} \sigma _{x} - \delta(y)\sigma _{z},  
\label{eq:Hamiltonian}
\end{align}
where $ m $ represents the atomic mass, $ \hat{p} _{\mu } =-i\hbar\nabla_{\mu } $ ($\mu =x,y $) is the canonical momentum along $\mu$-direction, $\hbar k_{r} $ is the recoil momentum, and $\Omega $ denotes the strength of the Raman coupling. Without loss of generality, we have assumed a two-dimensional (2D) condensate confined in a harmonic trapping potential $V_{trap}( x,y )  = m( \omega_{x} ^{2}  x^{2}+\omega_{y} ^{2} y^{2} ) /2$. The nonlinear interaction between atoms can be taken into account through the mean-field approximation $V_{int}= ( 1/2 ) \int d \mathbf{r} {\textstyle \sum_{\alpha \beta }}g_{\alpha \beta }n_{\alpha } n_{\beta }  $, where $g_{\alpha \beta }=4\pi  \hbar ^{2} a_{\alpha \beta } /m$ are the coupling constants in different spin channels related to the corresponding s-wave scattering lengths $ a_{\alpha \beta } $. In this work, we assume isotropic coupling constants $ g_{\alpha \beta } \equiv g$. The last term in Eq.~(\ref{eq:Hamiltonian}) corresponds to a spatially varying detuning $\delta  ( y ) =\eta k_{r} y$ that induces a synthetic magnetic field, with $\eta$ being the detuning gradient. Vortices appear when the detuning gradient exceeds a critical value $\eta > \eta _{c} $, whereas for $\eta < \eta _{c} $, the formation of quantized vortices is energetically inhibited, resulting in vortex-free condensates.

For simplicity, we focus on the dynamics within the zero-momentum phase, characterized by $\Omega > \Omega_c \equiv 4 E_r$, where $E_r = \hbar^{2} k_r^{2}/2m$ is the recoil energy. The order parameter for each component is parameterized as $\psi_{\alpha }   =\sqrt{n_{\alpha } ( \mathbf{r} )  } e^{i\phi _{\alpha }  ( \mathbf{r}  )  }$ \cite{BECbook}, with $n_{\alpha } $ and $\phi _{\alpha } $ representing the density and phase of the $\alpha$-th component. For ground states and low-frequency excitations, the relative phase of the two components is locked due to the large excitation gap, resulting in $\phi _{1} =\phi _{2}\equiv \phi  $ \cite{PhysRevA.86.063621}. Furthermore, when the emergent spin density $s_z=n_1-n_2$ is much smaller than the total density $n=n_1+n_2$, one can obtain a simple relation 
\begin{equation}
s_{z} = ( \frac{2\hbar ^{2}k_{r}  }{m\Omega } \nabla_{x} \phi + \frac{2\eta k_{r} }{\Omega }y  )n  
\label{eq:sz}
\end{equation}
which is intimately related the spin-momentum locking. Neglecting the influence of the quantum pressure in the kinetic energy, the hydrodynamic equations for BECs in the synthetic magnetic field can be derived as follows
\cite{PhysRevLett.120.183202,PhysRevA.108.053316} 
\begin{eqnarray}
&&	\frac{\partial n}{\partial t} +  \nabla\cdot ( n\mathbf{v }) =0, 
 \label{eq:HD.density}
\\
&&	\hbar \frac{\partial \phi }{\partial t} +  \frac{1}{2}m^{\ast } v_{x}^{2}+ \frac{1}{2} mv_{y}^{2} +V_{trap}+ng-\frac{\Omega }{2}=0
 \label{eq:HD.phase}
\end{eqnarray}
where the effective mass for the zero-momentum phase is given by $ m^{\ast } =m ( 1-\Omega _{c}/ \Omega   ) ^{-1} $. The physical velocity field 
\begin{equation}
\mathbf{v} =( \frac{\hbar }{m^{\ast } }\nabla_{x}\phi -\frac{\eta }{\hbar }\frac{\Omega _{c} }{\Omega } y, \frac{\hbar }{m} \nabla_{y} \phi ) 
\label{eq:velocity}
\end{equation}
includes the canonical velocity field and a novel term associated with the spin density. In the presence of a synthetic magnetic field, the spin contribution becomes nonzero, leading to a rigid-like rotational velocity field~\cite{PhysRevLett.118.145302,PhysRevLett.120.183202,PhysRevA.108.053316}.

\begin{figure}[t!]
\centering
\includegraphics[width=1\linewidth]{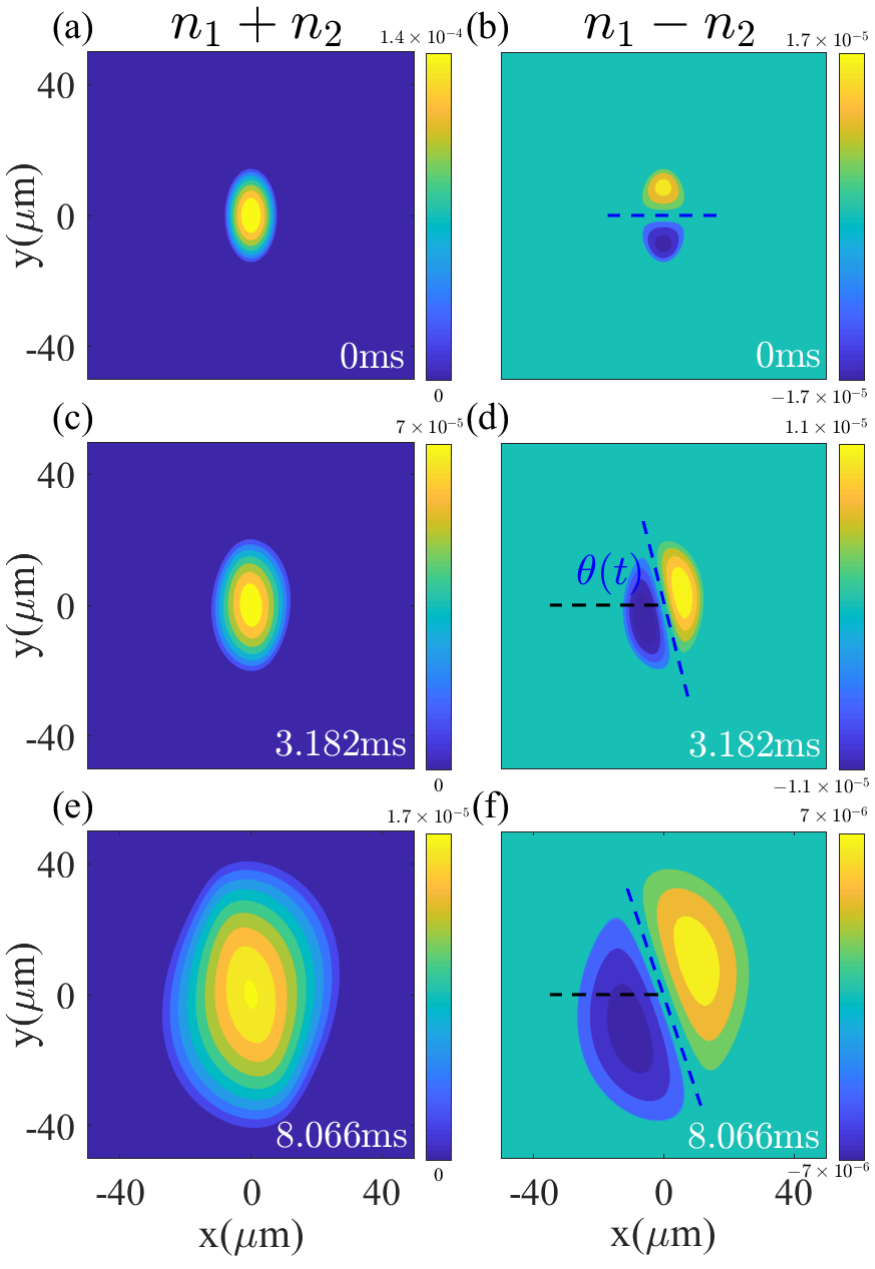}
\caption{Evolution of total density $ n$ and spin density $ s_{z}$ of a 2D BEC in a synthetic magnetic field at various expansion times $t$ (indicated in each panel). The results are obtained from GP simulations with Raman coupling $\Omega=6E_{r}$, detuning gradient $\eta=0.005E_{r}$, and anisotropic trapping frequencies $  ( \omega _{x},\omega _{y}    )=2\pi \times ( 50\sqrt{3} ,50 ) $Hz. The Thomas-Fermi radii at $t=0$ are $R_{x0} \approx 12.3\mu m$ and $R_{y0} \approx 21.5\mu m$, respectively. As shown in (d), the deflection angle of the spin density is $\theta(t)$, which is defined as a clockwise rotation angle. At $t=0$, $\theta(t)=0$. The dark blue and light yellow of the colorbar correspond to the density minimum and maximum, respectively. }
\label{fig:density}
\end{figure}

At equilibrium, the total density of a vortex-free condensate is well described by the Thomas-Fermi distribution, as shown in Fig.~\ref{fig:density}(a). For $\eta=0$, the condensate is unpolarized in the zero-momentum phase, implying a zero spin density. However, for finite $\eta$, the detuning gradient introduces a spin-dependent potential along the $y$-direction: $\frac{1}{2} m\omega_{y}^{2}y^2-\delta (y)\sigma _{z} =  \frac{1}{2} m\omega_{y}^{2}(y\pm y_{0})^{2}$ (up to a constant) with $y_{0} =\eta k_{r} / ( m\omega _{y}^{2}   ) $. As shown in Fig.~\ref{fig:density}(b), this results in a separation of the two spin components along $y$-direction. For small detuning gradient, the total density can be written as the parabolic form $n ( \mathbf{r} ,0  ) =n_{0}  ( 1-x^{2} /R_{x0}^{2}  -y^{2} /R_{y0}^{2}   )$ \cite{PhysRevLett.77.5315,PhysRevA.54.R1753,PhysLettA.10.1016/S0375-9601(97)00069-8} and the phase takes the form $\phi _{0}=\alpha _{0} xy  $. Here, $ n_{0} =2N/(\pi R_{x0}R_{y0}) $ is the peak density, $N$ is the number of atoms, and $R_{\mu 0} $ is the Thomas-Fermi radius along $\mu$-direction. According to Refs.~\cite{PhysRevLett.118.145302,PhysRevLett.120.183202}, the velocity field and spin density can be explicitly expressed as $\mathbf{v} = (  y( \hbar \alpha _{0}/m^{\ast } -\eta \Omega _{c}/  \hbar \Omega     ),    \hbar \alpha _{0} x /m   ) $ and $s_{z} =2\beta _{0}yn $, respectively, with $\alpha _{0} =2\eta k_{r}^{2} \omega _{x}^{2} / ( \Omega ( \omega _{x}^{2}m/m^{\ast } +  \omega _{y}^{2}  )    )$ and $\beta _{0}=k_{r} ( \eta +  \alpha _{0}\hbar ^{2} /m  )/\Omega $.

\section{Expansion dynamics in the absence of vortices}
Upon turning off the harmonic trap, the condensate starts to expand. Figure~\ref{fig:density} shows the typical evolution of the total density and spin density of a BEC in a synthetic rotational field after various expansion times obtained by numerically solving the GP equations. We note that the total density of the condensates can still be reasonably depicted by the Thomas-Fermi distribution after an expansion time up to $t\sim 4ms$. During the expansion, the two spin components rotate and separate, resulting in a rapid deflection of the spin density which is characterized by the angle $\theta(t)$ (as illustrated in Fig.~\ref{fig:density}(d)). After reaching a maximum value at the expansion time $t=3.182ms$, the deflection angle starts to decrease slowly. Unlike the spin density, the total density only begins to deviate significantly from the Thomas-Fermi distribution after a much longer expansion time.

The short-time expansion dynamics and the deflection of the spin density can be accurately predicted from the hydrodynamic theory. Since the total density maintains the Thomas-Fermi distribution for a short expansion time, it can be assumed to be
\begin{equation}
	n ( \mathbf{r} ,t  ) =n_{0} ( t  ) \left( 1-\frac{x^{2} }{R_{x}^{2} ( t ) }- \frac{y^{2} }{R_{y}^{2} ( t ) } \right),  
\label{eq:density}
\end{equation}
where $ n_{0}  ( t  ) =2N/ ( \pi R_{x} ( t  ) R_{y} ( t )   ) $ and $ R_{\mu }  ( t  ) $ are the peak density and Thomas-Fermi radii during the expansion. 

The phase of the expanding condensates takes the following ansatz 
\begin{equation}
	\phi ( t  ) =\frac{1}{2} \alpha _{x}  ( t  ) x^{2} + \frac{1}{2} \alpha _{y}  ( t  ) y^{2} +  \alpha ( t  ) xy. 
 \label{eq:phase}
\end{equation}
Substituting this into Eq.~(\ref{eq:velocity}), we find the velocity field becomes $v_{x}  ( t ) =\hbar /m^{\ast }  ( \alpha _{x} ( t  )x+\alpha( t  )y  )-\eta \Omega _{c}/ ( \hbar \Omega   ) y$, $v_{y}  ( t  ) =\hbar ( \alpha _{y} ( t  )y+\alpha  ( t  )x )/m  $. Substituting these expressions into the hydrodynamic Eq.~(\ref{eq:HD.density}), we obtain 
\begin{eqnarray}
 \alpha _{x} (t)&=&\frac{m^{\ast }\dot{b}_{x}(t)}{\hbar b_{x}(t)}  \\
 \alpha _{y} (t)&=&\frac{m\dot{b}_{y}(t)}{\hbar b_{y}(t)} \\
 \alpha(t) &=&\frac{2\eta k_{r}^{2} \omega _{x}^{2} b_{y}^{2}(t) }{ \Omega ( \frac{m}{m^{\ast }}\omega _{x}^{2}b_{y}^{2}(t)  +  \omega _{y}^{2} b_{x}^{2}(t))}
\end{eqnarray}
where we have introduced $ b_{\mu }  ( t  )=R_{\mu}(t)/R_{\mu 0}$ as the expansion parameter along $\mu$-direction. After turning off the trapping potential ($V_{trap}=0$), using the hydrodynamic Eq.~(\ref{eq:HD.phase}) and neglecting higher-order terms of $\eta  $, we find that the expansion parameter $b_{\mu }(t)$ satisfies the following equation of motion
\begin{equation}
\ddot{b} _{\mu }  ( t  ) - \frac{\tilde{\omega }_{\mu }^{2}  }{b_{\mu }  ( t  )b_{x}  ( t  ) b_{y}  ( t  )  } =0,  
\label{eq:differential}
\end{equation}
with $ \tilde{\omega }_{x} =\omega _{x} \sqrt{m/m^{\ast } }$, $\tilde{\omega }_{y} =\omega _{y} $. The initial conditions can be determined from the properties of the system at equilibrium. It is easy to find $ \alpha _{x}  ( 0  ) =\alpha _{y}  ( 0  )=0$, $\alpha ( 0  )=\alpha _{0} $, $ b_{\mu }  ( 0  )=1$ and $ \dot{b} _{\mu }  ( 0  ) =0$.

\begin{figure}[t!]
\centering
\includegraphics[width=1\linewidth]{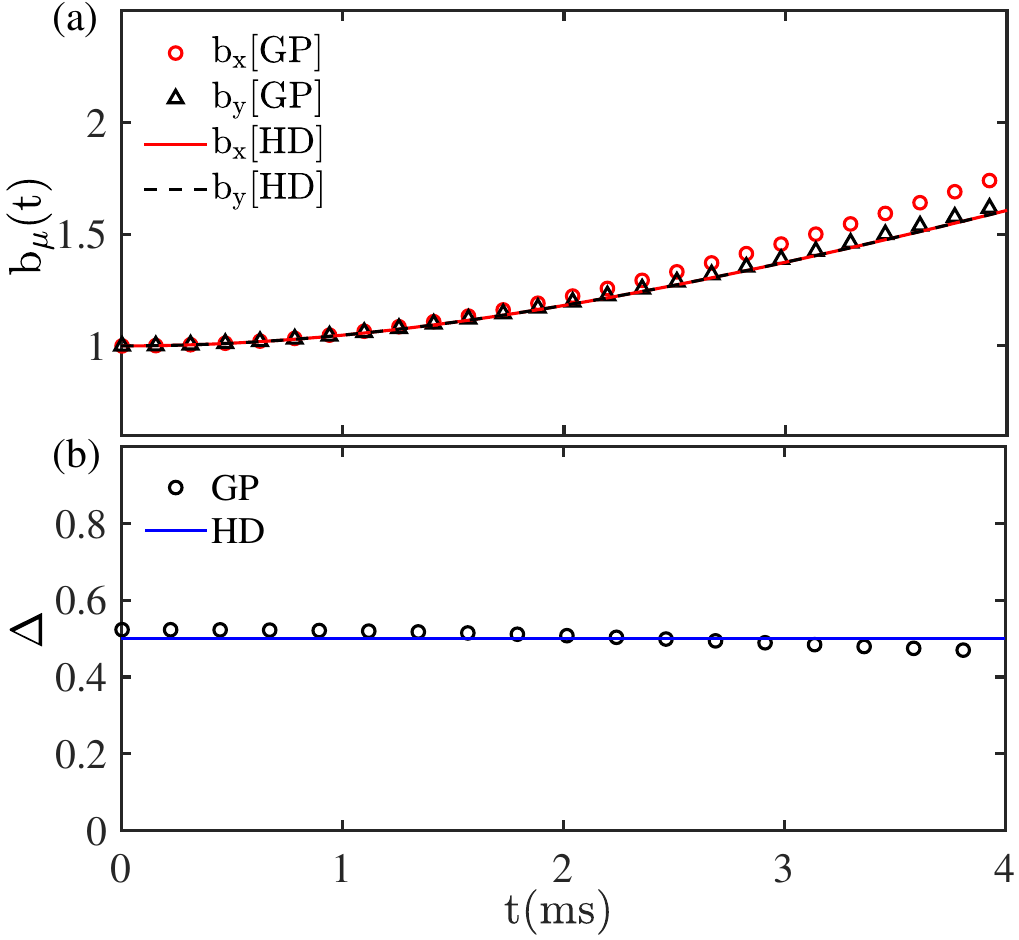}
\caption{Time evolution of the expansion parameters $b_{\mu}=R_{\mu } ( t  ) /R_{\mu 0} $ (a) and the deformation parameter $\Delta$ (b) after turning off the harmonic trap. The solid and dashed lines are the hydrodynamic predictions and the symbols are the results obtained from the GP simulation. The system parameters are the same as that for Fig.~\ref{fig:density}.}
\label{fig:radius}
\end{figure} 

A particularly interesting scenario occurs when the trapping frequencies satisfy $ \sqrt{m/m^{\ast } } \omega _{x} =\omega _{y} \equiv \omega _{0} $, i.e., when the two dipole oscillation frequencies are degenerate. Under this condition, the expansion parameters are equal and can be analytically obtained as $b_{x}  ( t  ) = b_{y}  ( t  ) =\sqrt{1+\omega _{0}^{2} t^{2} } $. In the following, we will focus on this parameter regime. 

To better understand the time evolution of the condensate density during the expansion, we introduce the deformation parameter which is defined as:
\begin{equation}
	\Delta(t)  =\frac{ \left\langle y^{2}-x^{2}    \right\rangle }{\left \langle y^{2} +  x^{2} \right \rangle } =\frac{R_{y}^{2}  ( t  )- R_{x}^{2}  ( t  )  }{R_{y}^{2}  ( t  )+   R_{x}^{2}  ( t  )}. 
 \label{eq:deformation}
\end{equation} 
For the specific trap frequencies $(\omega_{x},\omega_{y})=2\pi \times (50\sqrt{3},50) $Hz, the aspect ratio $R_{x}  ( t ) / R_{y}  ( t  )=1/\sqrt{3} $ and the deformation parameter $\Delta  =1/2$. As shown in Fig.~\ref{fig:radius}, the time evolution of the expansion parameters $b_\mu(t)$ and the deformation parameter $\Delta(t)$ from the hydrodynamic theory predictions agree very well with those obtained from the GP simulations. Note that in the GP simulation, we numerically compute the Thomas-Fermi radii during the expansion using the related $ R_{\mu }  ( t  )=\sqrt{12\left \langle x_{\mu }^{2}  \right \rangle }$ and subsequently calculate the expansion parameters and deformation parameter.
 
Substituting the expression for the total phase (Eq.~(\ref{eq:phase})) into the spin density expression $s_{z}(t)$ (Eq.~\ref{eq:sz}), we find the following explicit form
 \begin{align}
 \frac{s_{z}(t)  }{n( t ) } = \frac{2\hbar^{2}k_{r}}{m\Omega }\alpha _{x}(t)x   + 2(\frac{k_{r}\eta  }{\Omega}+\frac{\hbar^{2}k_{r}}{m\Omega }\alpha(t))y.
 \label{eq:spin}
\end{align} 
Since $\alpha(t)\propto \eta$, in the absence of a position-dependent detuning ($\eta=0$), the spin density is zero in the ground state and it becomes $x$-dependent in the expansion, resulting in a spin separation along $x$~\cite{NewJphys.10.1088/1367-2630/aa7e8c}. When $\eta \neq 0$, the two spin components are separated along $y$ even at equilibrium because $\alpha(t=0)\neq 0$. After the trapping potential is turned off, the spin density starts to develop an $x$-dependent term. The presence of both $x$- and $y$-dependent terms in Eq.~(\ref{eq:spin}) implies a rotation of the spin density characterized by the deflection angle. This angle is related to the ratio of the $x$- and $y$-dependent terms in Eq.~(\ref{eq:spin}). Thus, it can be analytically obtained as
 \begin{equation}
\theta ( t  ) =\frac{\pi }{2}-\arctan(\frac{\eta }{2\hbar }(2-\frac{\Omega _{c} }{\Omega })\frac{1+\omega _{0}^{2}t^{2}  }{\omega _{0}^{2}t }).    
 \label{eq:theta}
\end{equation}   

\begin{figure}[t!]
\centering
\includegraphics[width=1\linewidth]{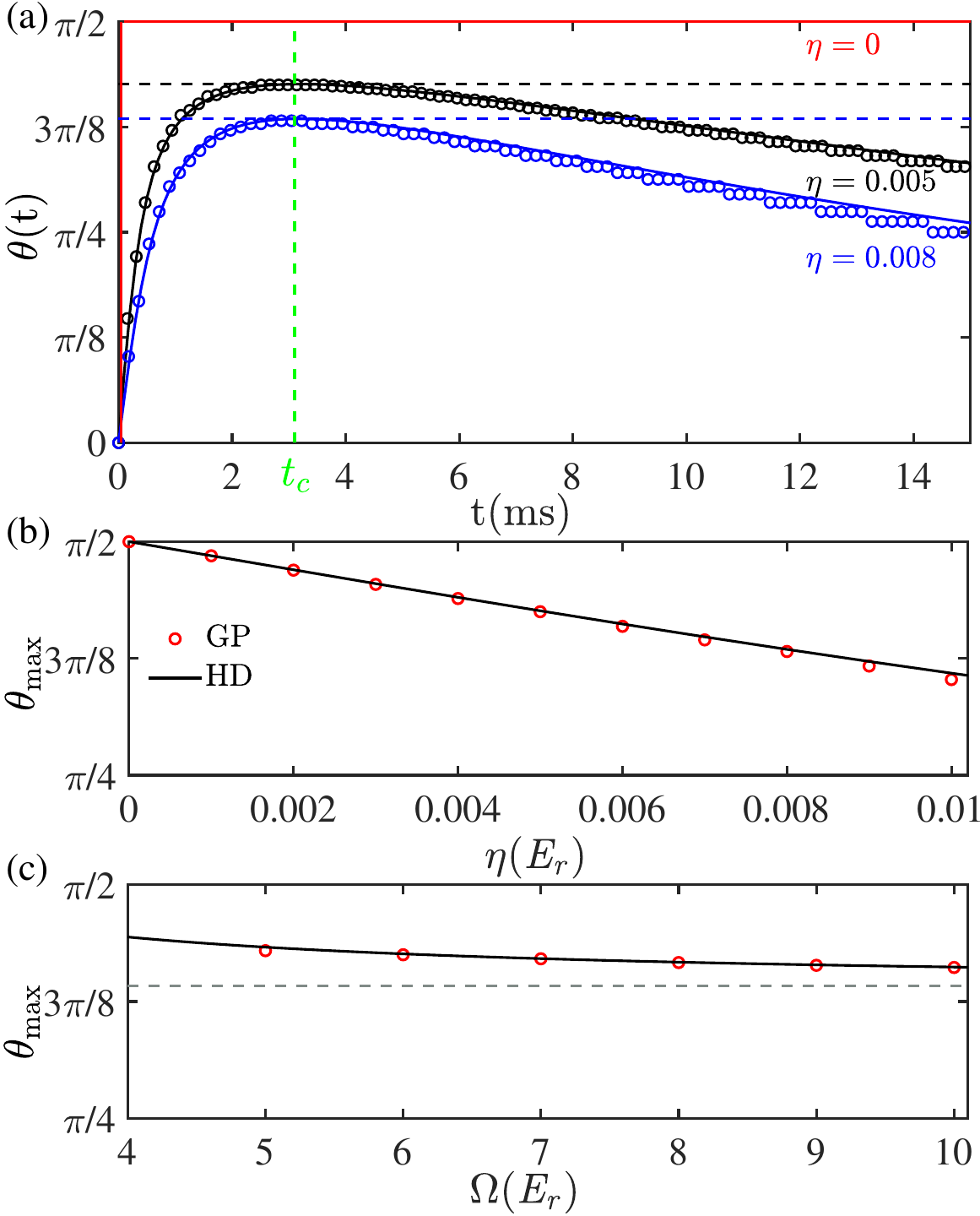}
\caption{(a) The deflection angle of the spin density as a function of the expansion time for different detuning gradients $\eta=0$ (red), $0.005E_r$ (black), and $0.008E_r$ (blue). The maximum deflection angle as a function of $\eta$ and Raman coupling strength $\Omega$ is shown in (b) and (c), respectively. In all the panels, the solid lines are the results obtained from the hydrodynamic theory, while the symbols are obtained by numerically solving the GP equations. The black and blue dashed horizontal lines in panel (a) mark the maximum deflection angle $\theta_{max}$ which occurs $t_{c}=1/\omega_{0}$ (indicated by the green dashed vertical line). In (a), the step-wise behavior of the circles at longer expansion times is related to the numerical resolution in the GP simulation. The dashed gray line in (c) corresponds to the asymptotic limit $\Omega \to \infty $. }
\label{fig:theta}
\end{figure}

\begin{figure}[t!]
\centering
\includegraphics[width=1\linewidth]{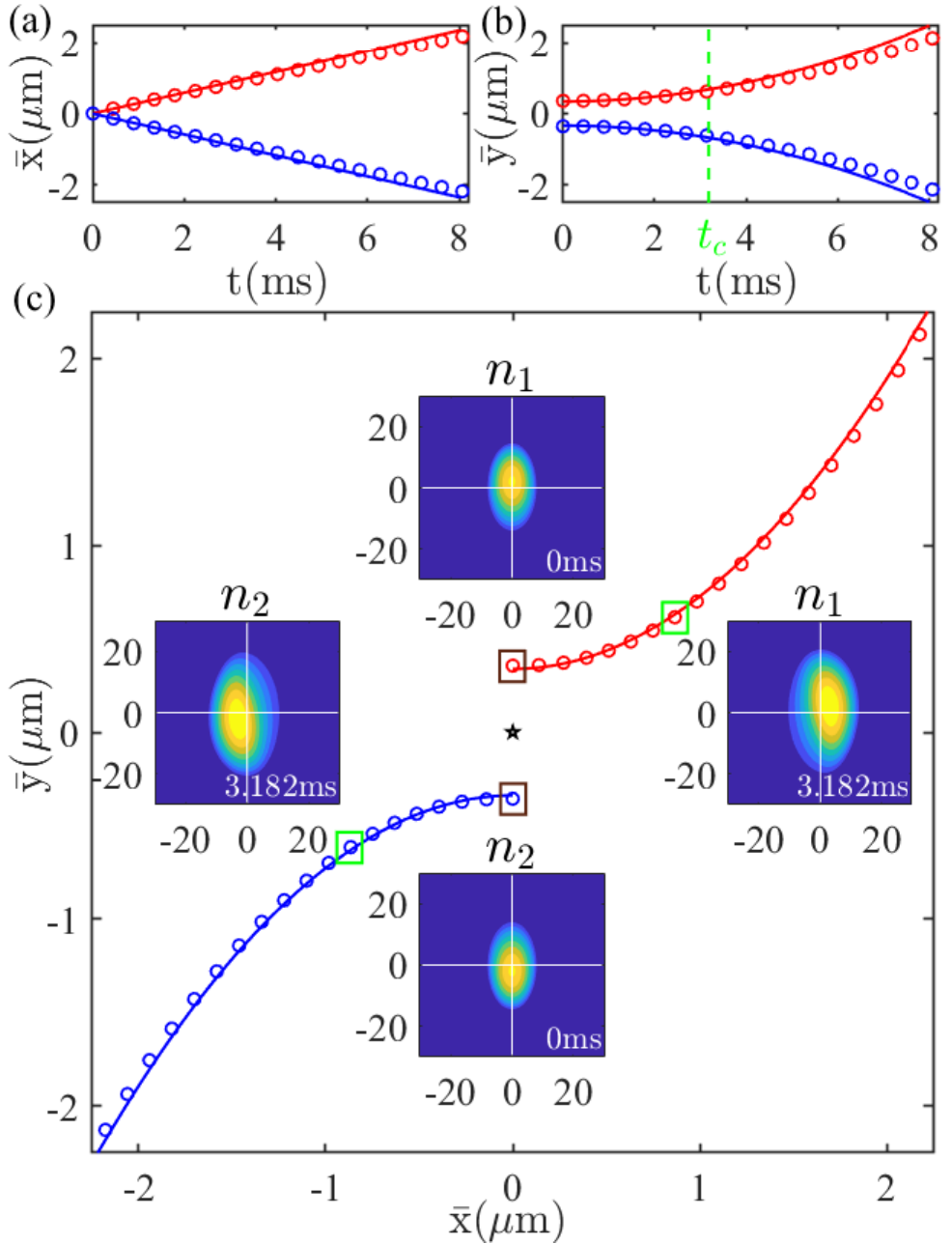}
\caption{The center-of-mass motion of the two spin states after switching off the external trapping potential, where red (blue) circles and solid lines represent the center-of-mass position of the spin up (down) state calculated by GP and hydrodynamic equations, respectively. (a) and (b) show the center-of-mass position versus the expansion time along $x$ and $y$, respectively, with $\eta=0.005E_{r}$, $\Omega=6E_{r}$. The trajectory of $(\bar{x}_{n_{\alpha } }, \bar{y}_{n_{\alpha } })$ obtained from (a) and (b) is shown in (c). The inset panels in (c) show the 2D densities $n_{\alpha } (x,y)$ of the two spin components at an expansion time $t = 0$ and $t = 3.182ms$, with the corresponding center-of-mass position indicated by the brown and green squares, respectively. The black star is the center-of-mass position of total density, which remains unchanged during the expansion.}    
\label{fig:position}
\end{figure}

The deflection angle predicted from the hydrodynamic theory (Eq.~(\ref{eq:theta})) and calculated by numerically solving the coupled GP equations are shown in Fig.~\ref{fig:theta}. Without a synthetic magnetic field, the two spin states are overlapped and they are located at the minimum of harmonic trap at $t=0$. Once released from the trap, the two components only separate along $x$ and thus the defined deflection angle for the expanding spin density becomes $\pi /2 $ (red line in Fig.~\ref{fig:theta}(a)). However, in the presence of a synthetic magnetic field, the two components do not overlap and are separated along $y$ even at $t=0$. Once released from the trap, the deflection angle increases from zero to the peak value at $t_{c} = 1/\omega_{0} $, and then it slowly decreases. The expression for the maximum deflection angle $\theta _{max}$ can be easily obtained from the hydrodynamic theory
\begin{equation}
 	\theta _{max} =\frac{\pi }{2} - \arctan \frac{\eta }{\hbar \omega _{0} } ( 2-  \frac{ \Omega_{c} }{\Omega}   ). 
   \label{eq:maxtheta}
\end{equation} Figure~\ref{fig:theta}(b) illustrates the quantitative dependence of $\theta _{max}$ on the detuning gradient $\eta$ for Raman coupling strength $\Omega=6E_{r}$. The maximum deflection angle decreases as the detuning gradient increases. Fig.~\ref{fig:theta}(c) shows $\theta _{max}$ versus $\Omega$ for a given detuning gradient $\eta=0.005E_{r}$. We find that as $\Omega \to \infty $, $\theta _{max} \to
\pi/2-\arctan(2\eta/\hbar\omega_0)$ (indicated by the dashed grey line in Fig.~\ref{fig:theta}(c)). Conversely, as $\Omega \to \Omega_{c} $, $\theta _{max} \to \pi/2-\arctan(\eta/\hbar\omega_0)$.

Thus far, we have focused on the evolution of total density and spin density in the expansion dynamics. We may gain further insights by studying the center-of-mass motion of each component. The density of the individual components can be easily obtained as $n_{\alpha }= ( n\pm s_{z}   )/2  $. From this, we can derive the time evolution of the center-of-mass position of the two spin states. Using the definitions $\bar{x}_{n_{\alpha } } =\int  ( xn_{\alpha } ) d\mathbf{r}  $, $\bar{y}_{n_{\alpha } } =\int  ( yn_{\alpha }  ) d\mathbf{r}  $, we find
\begin{eqnarray}
  \bar{x} _{n_{1,2} }  ( t  ) &=&\pm \frac{N\hbar k_{r} }{6(\Omega-\Omega_{c})  }R_{x0}^{2}\omega _{0}^{2}t,
   \label{eq:x-motion} \\
   \bar{y} _{n_{1,2} }  ( t  ) &=& \pm \frac{Nk_{r}\eta  }{12\Omega  }(1+\frac{m^{\ast } }{m}) R_{y0}^{2}(1+\omega _{0}^{2}t^{2}). 
   \label{eq:y-motion}
\end{eqnarray}
From these results, we see again that the two components are overlapped along $x$ at $t=0$ and then they are separated with the separation increasing linearly in time (Fig.~\ref{fig:position}(a)). On the other hand, the two components are separated along $y$ even at $t=0$ and then their separation increases quadratically in time (Fig.~\ref{fig:position}(b)). Consequently, the trajectory of the two expanding components in the $x-y$ plane follows a parabolic curve (Fig.~\ref{fig:position}(c)). The deflection angle $\theta(t)$ for the spin density is related to the slope of the line connecting the origin $(0, 0)$ and the position at time $t$, i.e., $(\bar{x}_{n_\alpha}(t), \bar{y}_{n_\alpha}(t))$. Apparently, the deflection angle becomes maximum when the line is tangential to the trajectory which happens at $t_c= 1/\omega_0$.

\begin{figure}[t!]
\centering
\includegraphics[width=1\linewidth]{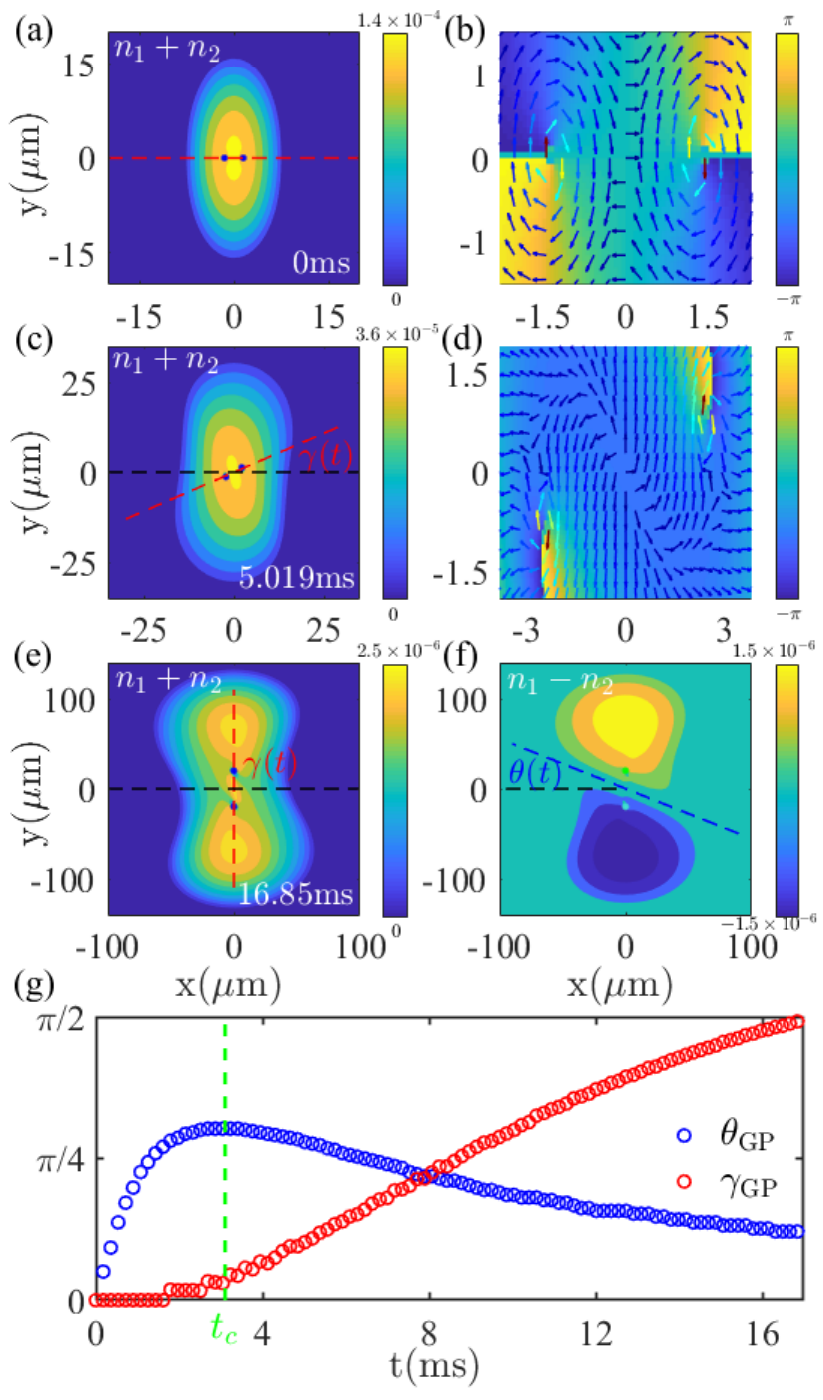}
\caption{Expansion dynamics in the presence of vortices when $\eta>\eta_{c}$. (a) and (c) show the expanding total density at $t=0$ and $t=5.019ms$, respectively, with the corresponding velocity field (arrows) near the center of vortices shown in (b) and (d). The color of the arrows reflects the magnitude of the velocity field, where red arrows represent large velocities and blue ones represent small velocities. The color map of the background in (b) and (d) illustrates the phase distributions of order parameters, where the phase changes by $2\pi$ after completing a closed loop around a vortex. $\gamma (t) $ in (c) and (e) is the rotation angle of the two vortices, defined positive counterclockwise. (e) and (f) respectively show the total density and spin density after a long expansion time $ t=16.85$ms. The GP simulation results for the rotation angle of the vortices and the deflection angle of the spin density are shown in (g) with $\eta=0.0013E_{r}$ and $\Omega=6E_{r}$.}    
\label{fig:vortex}
\end{figure}   

\section{Expansion dynamics in the presence of vortices}
In this section, we consider the expansion dynamics in the presence of vortices for $\eta > \eta _{c} $. Because of the emergence of vortices, the density of the condensate is not smooth and one cannot apply the hydrodynamic theory. Therefore, we focus on the numerical simulation of the GP equations to demonstrate the expansion dynamics.

The number of the vortices appearing in the condensate depends on the detuning gradient. With $\eta=0.0013E_{r}$ and $\Omega=6E_{r}$, through imaginary time evolution of the coupled GP equations, we find that the ground state contains two vortices located on the $x$-axis and near the origin (see Fig.~\ref{fig:vortex}(a)). The corresponding velocity field and the phase around the vortex cores are illustrated in Fig.~\ref{fig:vortex}(b), showing a negative circulation $\mathcal{C} =\oint \mathbf{v}\cdot d\mathbf{l}  $ as it is opposite to the direction of the rigid-like rotational velocity field of the background condensate~\cite{PhysRevA.84.063604}. After turning off the external trapping potential, in addition to the separation and rotation of the two components, we find that the two vortices begin to rotate counterclockwise which is the same as the direction of the rotational velocity field (Fig.~\ref{fig:vortex}(c)). However, the corresponding velocity field near the vortex cores remains clockwise (Fig.~\ref{fig:vortex}(d)). After a long-time expansion of $t=16.85$ms, the total density and spin density are shown in Fig.~\ref{fig:vortex}(e) and (f), respectively. The deflection angle of the spin density $\theta(t)$ reaches the maximum value at $t=t_{c}$, similar to the case when $\eta < \eta _{c} $. However, as shown in Fig.~\ref{fig:vortex}(g), the rotation angle $\gamma(t)$ of the two vortices increases linearly with time when the expansion time $t>t_{c}$. The rotation angle of the two vortices is predominately related to the rotation of the total density due to the synthetic rotational field while the deflection of the spin density is mainly related to the center-of-mass motion of the two spin components.     

\section{Conclusion}
In summary, we have investigated the expansion dynamics of SO-coupled BECs in the presence of a position-dependent detuning. We demonstrate how the generated synthetic magnetic field and rigid-like rotational velocity field lead to the interesting phenomena of separation and rotation of the two components. By developing a hydrodynamic theory, we have analytically determined the maximum deflection angle that occurs at a characteristic time $t_c= 1/\omega_0$. Our hydrodynamic theory predictions align very well with numerical GP simulation results. We also have extended the study to the expansion dynamics after the formation of vortices when the detuning gradient exceeds a critical value. In addition to the rotation of the spin density, the vortices undergo additional rotation dynamics, with the rotation angle increasing approximately linearly for expansion time $t>t_{c}$. 

Our finding offers valuable insights into the effects of the synthetic magnetic field on the expansion dynamics. It may catalyze further experimental and theoretical explorations on the non-equilibrium dynamics of SO-coupled BECs in the presence of synthetic magnetic fields and rigid-like rotational velocity fields. The fact that the rotation of the spin density depends on the detuning gradient suggests that this system may be a promising candidate for the development of a magnetic field gradiometer. 

\begin{acknowledgments}
J.W., H.L., and Y.L. are supported by National Natural Science Foundation of China (Nos. 11774093, 12074120), Natural Science Foundation of Shanghai (No. 23ZR118700), Innovation Program of Shanghai Municipal Education Commission (No. 202101070008E00099) and Program of Chongqing Natural Science Foundation (CSTB2022NSCQ-MSX0585).
C.H.L. acknowledges support from Quantum Science Center (QSC), a Department of Energy (DOE) quantum information science (QIS) research center, and related earlier experimental work supported in part by NSF Grant No. PHY-2012185.
\end{acknowledgments}

%\nocite{*}

\bibliography{ref}% Produces the bibliography via BibTeX.

\end{document}